\documentclass[12pt]{iopart}

\usepackage{iopams}  

\usepackage{graphicx}
\usepackage{dsfont}
\usepackage{braket}
\usepackage{color}
\usepackage{soul}
\usepackage[bookmarks=true, colorlinks]{hyperref}
\usepackage{lineno}

\newcommand{\tab}[1]{table.~\ref{#1}}
\newcommand{\fig}[1]{figure~\ref{#1}}
\newcommand{\eq}[1]{(\ref{#1})}
\newcommand{\Sec}[1]{section~\ref{#1}}

\newcommand{\I}{{\rm i}}
\newcommand{\subtiny}[3]{\ensuremath{_{\hspace{#1 pt}\protect\raisebox{#2 pt}{\tiny{$ #3$}}}}}

\newcommand*{\citen}[1]{reference~\cite{#1}}

\begin{document}

\title[Speeding-up the decision making of a learning agent]{Speeding-up the decision making of a learning agent using an ion trap quantum processor}

\author{Th Sriarunothai$^1$, S W{\"o}lk$^{1,5}$, G S Giri$^{1,6}$, N Friis$^{2,7}$, V Dunjko$^{2,3,8}$, H J Briegel$^{2,4}$ and Ch Wunderlich$^1$}

\address{$^1$ Department of Physics, School of Science and Technology, University of Siegen, 57068 Siegen, Germany}
\address{$^2$ Institute for Theoretical Physics, University of Innsbruck, Technikerstra{\ss}e 21a, 6020 Innsbruck, Austria}
\address{$^3$ Max Planck Institute of Quantum Optics, Garching 85748, Germany}
\address{$^4$ Department of Philosophy, University of Konstanz, 78457 Konstanz, Germany}
\address{$^5$ Present address: Institute for Theoretical Physics, University of Innsbruck, Technikerstra{\ss}e 21a, 6020 Innsbruck, Austria}
\address{$^6$ Present address: Institut f{\"u}r Experimentalphysik, Heinrich-Heine-Universit{\"a}t D{\"u}sseldorf, 40225 D{\"u}sseldorf, Germany}
\address{$^7$ Present address: Institute for Quantum Optics and Quantum Information, Austrian Academy of Sciences, Boltzmanngasse 3, 1090 Vienna, Austria}
\address{$^8$ Present address: LIACS, Leiden University, Niels Bohrweg 1, 2333 CA Leiden, The Netherlands}

\ead{christof.wunderlich@uni-siegen.de}


\begin{abstract}
We report a proof-of-principle experimental demonstration of the quantum speed-up for learning agents utilizing a small-scale quantum information processor based on radiofrequency-driven trapped ions. The decision-making process of a quantum learning agent within the projective simulation paradigm for machine learning is implemented in a system of two qubits. The latter are realized using hyperfine states of two frequency-addressed atomic ions exposed to a static magnetic field gradient. We show that the deliberation time of this quantum learning agent is quadratically improved with respect to comparable classical learning agents. The performance of this quantum-enhanced learning agent highlights the potential of scalable quantum processors taking advantage of machine learning.
\end{abstract}

%
\vspace{2pc}

%
%
%
%


\section{Introduction\label{sec:introduction}}

The past decade has seen the parallel advance of two research areas \textemdash\ quantum computation~\cite{Nielsen2000} and artificial intelligence~\cite{Russell2003} \textemdash\ from abstract theory to practical applications and commercial use. Quantum computers, operating on the basis of information coherently encoded in superpositions of states that could be considered classical bit values, hold the promise of exploiting quantum advantages to outperform classical algorithms, e.g., for searching databases~\cite{Grover1996}, factoring numbers~\cite{Shor1994}, or even for precise parameter estimation {with quantum metrology}~\cite{Giovanetti2004,Friis2017}. At the same time, artificial intelligence and machine learning have become integral parts of modern automated devices using classical processors~\cite{Lim2017,Silver2016,Mnih2015,Schaeffer1518}. Despite this seemingly simultaneous emergence and promise to shape future technological developments, the overlap between these areas still offers a number of unexplored problems~\cite{Biamonte2017,DunjkoReview2017}. It is hence of fundamental and practical interest to determine how quantum information processing and autonomously learning machines can mutually benefit from each other.

\begin{figure}[h]
	\centering
	(a)\includegraphics[width=0.5\textwidth]{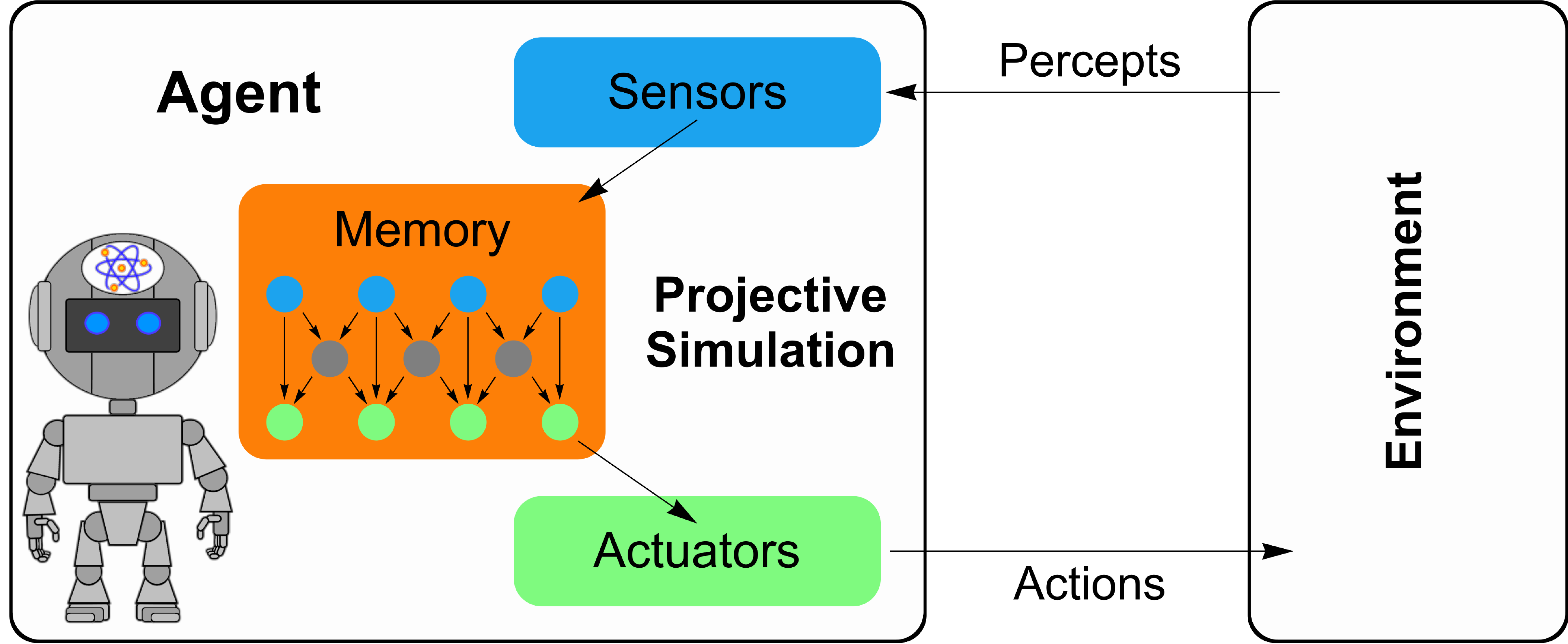}
	(b)\includegraphics[width=0.5\textwidth]{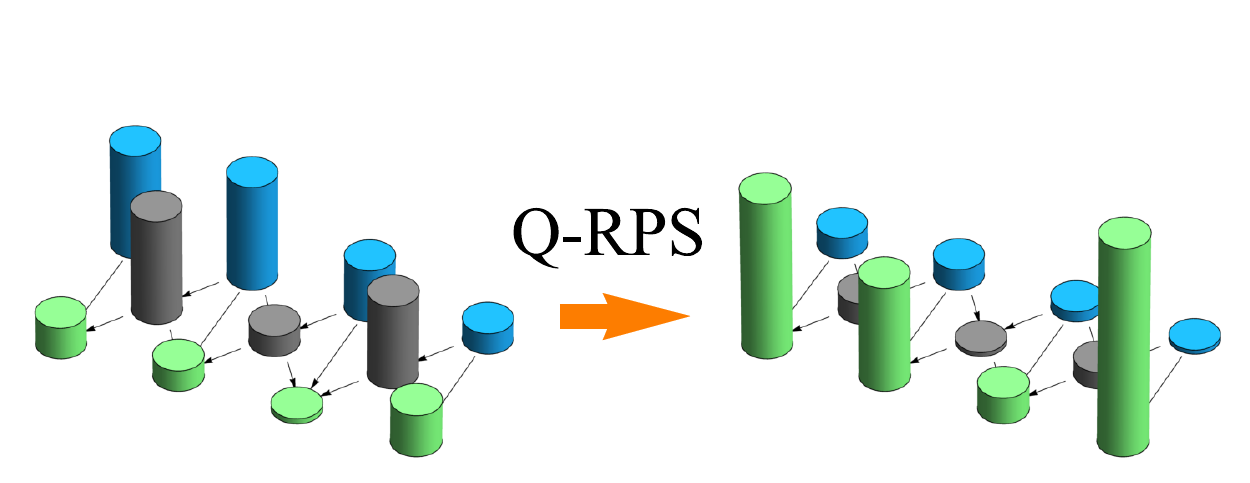}
	\caption{Learning agent \& quantum reflecting projective simulation (Q-RPS). (a) Learning agents receive perceptual input (``percepts'') from and act on the environment. The projective simulation (PS) decision-making process draws from the agent's memory and can be modeled as a random walk in a clip network, which, in turn, is represented by a stochastic matrix $P$. 
	Here the clips represent the elementary patches of episodic memory of prior experiences.
(b) Q-RPS agents enhance the relative probability of (desired) actions (green columns) compared to other clips (grey) that may include undesired actions or percepts (blue) within the stationary distribution of $P$ before sampling, achieving a quadratic speed-up {with respect} to classical RPS agents.}
	\label{fig:la-concept}
\end{figure}

Within the area of artificial intelligence, a central component of modern applications is the learning paradigm of an agent interacting with an environment~\cite{Sutton1998,Russell2003,Briegel2012}
illustrated in \fig{fig:la-concept}~(a), which is usually formalized as so-called reinforcement learning. This entails receiving perceptual input and being able to react to it in different ways. The learning aspect is manifest in the reinforcement of the connections between the inputs and actions, where the correct association is (often implicitly) specified by a reward mechanism, which may be external to the agent. In this very general context, an approach to explore the intersection of quantum computing and artificial intelligence is to equip autonomous learning agents with quantum processors for their deliberation procedure\footnote{Other approaches that we will not further discuss here concern among others models where internal processes are sped up by annealing processes~\cite{Crawford2016,Neukart2017}; or where the environment, and the agent's interaction with it may be of quantum mechanical nature as well~\cite{Dunjko2016,Dunjko2017,Alvarez-Rodriguez2017}.}.
That is, an agent chooses its reactions to perceptual input by way of quantum algorithms or quantum random walks. The agent's learning speed can then be quantified in terms of the average number of interactions with the environment until targeted behavior (reactions triggering a reward) is reproduced by the agent with a desired efficiency. This learning speed cannot generically be improved by incorporating quantum technologies into the agent's design~\cite{Dunjko2016}.

However, a recent model~\cite{Paparo2014} for learning agents based on projective simulation (PS)~\cite{Briegel2012} allows for a speed-up in the agent's deliberation time during each individual interaction. Theoretical work has shown that such a quantum improvement in the reaction speed should be possible within the reflecting projective simulation (RPS) variant of PS~\cite{Paparo2014}. There, the desired actions of the agent are chosen according to a probability distribution that can be modified during the learning process. This is of particular relevance to adapt to rapidly changing environments~\cite{Paparo2014}, as we shall elaborate on in the next section. For this task, the deliberation time of classical RPS agents is proportional to the quantities $1/\delta$ and $1/\epsilon${, where $\delta$ represents a spectral gap of a Markov chain and $\epsilon$ represents the probability to sample an action in a probability distribution}. These characterize the time needed to generate the specified distribution in the agent's internal memory and the time to sample a suitable (e.g., rewarded rather than an unrewarded) action from it, respectively. A quantum RPS (Q-RPS) agent, in contrast, is able to obtain such an action quadratically faster, i.e., within a time of the order $1/\sqrt{\delta\epsilon}$ as is shown in the next section.

Here, we report on the first proof-of-principle experimental demonstration of  a quantum-enhanced reinforcement learning system, complementing recent experimental work in the context of (un)supervised learning~\cite{Riste2017,Zhaokai2015,Cai2015}.
We implement the deliberation process of an RPS learning agent in a system of two qubits that are encoded in the energy levels of one trapped
atomic ion each. Within experimental uncertainties, our results confirm the agent's action output according to the desired distributions and within deliberation times that are quadratically improved with respect to comparable classical agents. This laboratory demonstration of speeding up a learning agent's deliberation process can be seen as the first experiment combining novel concepts from machine learning with the potential of ion trap quantum computers where complete quantum algorithms have been demonstrated~\cite{Hanneke2010,Monz2016,Piltz2016,Debnath2016}  and feasible concepts for scaling up~\cite{Kielpinski2002,Monroe2014,Lekitsch2017} are vigorously pursued.


\section{Theoretical Framework of RPS\label{sec:theory part}}

A generic picture for modeling autonomous learning scenarios is that of repeated rounds of interaction between an agent and its environment. In each round the agent receives perceptual input (``percepts") from the environment, processes the input using an internal deliberation mechanism, and finally acts upon (or reacts to) the environment, i.e., performs an ``action"~\cite{Briegel2012}. Depending on the reward system in place and the given percept, such actions may be rewarded or not, which leads the agent to update its deliberation process, the agent learns.

Within the projective simulation (PS)~\cite{Briegel2012} paradigm for learning agents, the decision-making procedure is cast as a (physically motivated) stochastic diffusion process within an episodic compositional memory, that is, a (classical or quantum) random walk in a representation of the agent's memory containing the interaction history. One may think of the {episodic compositional memory} as a network of clips that can correspond to remembered percepts, remembered actions, or combinations thereof. 
That is, the clips represent the elementary patches of episodic memory.
Mathematically, this clip network is described by a stochastic matrix (defining a Markov chain) $P=(p_{ij})$, where the $p_{ij}$ with $0\leq p_{ij}\leq1$ and $\sum_{i}p_{ij}=1$ represent transition probabilities between the clips labeled $i$ and $j$ with $i,j\in\{1,2,\ldots,N\}$. The learning process is implemented through an update of the $N\times N$ matrix $P$, which, in turn, serves as a basis for the random walks in the clip network. Different types of PS agents vary in their deliberation mechanisms, update rules, and other specifications.

In particular, one may distinguish between PS agents based on ``hitting" and ``mixing". For the former type of PS agent, a random walk could, for instance, start from a clip $c_{1}$ called by the initially received percept. The first ``step" of the random walk then corresponds to a transition to clips $c_{j}$ with probabilities $p_{1j}$. The agent then samples from the resulting distribution $\{p_{1j}\}_{j}$. If such a sample provides an action, for instance, if the clip $c_{k}$ is ``hit", this action is selected as output, otherwise the walk continues on from the clip $c_{k}$. An advanced variant of the PS model based on ``mixing" is reflecting projective simulation (RPS)~\cite{Paparo2014}. There, the Markov chain is first ``mixed", that is, an appropriate number\footnote{The mixing time depends on the spectral gap $\delta$ of the Markov chain $P$, i.e., the difference between the two largest eigenvalues of $P$~\protect\cite{Paparo2014}.}
of steps are applied until the stationary distribution is attained (approximately), before a sample is taken. This, or other implementations of random walks in the clip network provide the basis for the PS framework for learning. The classical PS framework can be used to solve standard textbook problems in reinforcement learning~\cite{Mautner2015,Melnikov2014,Makmal2016}, and has recently been applied in advanced robotics~\cite{Hangl2016}, adaptive quantum computation~\cite{Tiersch2015}, as well as in the machine-generated design of quantum experiments~\cite{Melnikov2018}.

Here, we focus on RPS agents, where the deliberation process based on mixing allows for a speed-up of Q-RPS agents {with respect} to their classical counterparts~\cite{Paparo2014}. In contrast to basic hitting-based PS agents, the clip network of RPS agents is structured into several sub-networks, one for each percept clip, and each with its own stochastic matrix $P$. In addition to being stochastic, these matrices specify Markov chains which are ergodic~\cite{Paparo2014}, which ensures that the Markov chain in question has a unique stationary distribution, i.e., a unique eigenvector $\pmb\alpha$ with eigenvalue $+1$, $P\pmb{\alpha}=\pmb{\alpha}$. Starting from any initial state, continued application of $P$ (or its equivalent in the quantized version) mixes the Markov chain, leaving the system in the stationary state.

As part of their deliberation process, RPS agents generate stationary distributions over their clip space, as specified by $P$, which is updated as the agent learns.
These distributions have support over the whole sub-network clip space, and additional specifiers -- flags --  are used to ensure an output from a desired sub-set of clips. For instance, standard agents are presumed to output actions only, in which case only the actions are flagged using standard emoticons \cite{Briegel2012}.
This ensures that an action will be output, while maintaining the relative probabilities of the
actions. 
Put simply, flags provide a mechanism that can be used as a short-term memory, or to mark actions, to (temporarily) store additional information about the clip network besides that contained in the Markov chain. The same mechanism of flags can also be used to eliminate iterated attempts of actions which did not yield rewards in recent time-steps. This leads to a more efficient exploration of correct behavior.

In the quantum version of RPS, each clip $c_{i}$ is represented by a basis vector $\ket{i}$ in a Hilbert space $\mathcal{H}$. In the most general case, the mixing process is then realized by a diffusion process on two copies of the original Hilbert space. On the doubled space $\mathcal{H}\otimes\mathcal{H}$ a unitary operator $W(P)$ (called the Szegedy walk operator~\cite{Szegedy2004,Magniez2011}) and a quantum state $\ket{\alpha^{\prime}}$ with $W(P)\ket{\alpha^{\prime}}=\ket{\alpha^{\prime}}$ take the roles of the classical objects $P$ and $\pmb\alpha$. Both $W(P)$ and $\ket{\alpha^{\prime}}$ depend on a set of unitaries $U_{i}$ on $\mathcal{H}$ that act as $U_{i}\ket{0}=\sum_{j}\sqrt{p_{ij}}\ket{j}$ for some reference state $\ket{0}\in\mathcal{H}$ . The more intricate construction of $W(P)$ is given in detail in {\citen{Dunjko2015}}.

The feature of the quantum implementation of RPS that is crucial for us here is an amplitude amplification similar to Grover's algorithm~\cite{Grover1996}, which incorporates the mixing of the Markov chain and allows outputting flagged actions after an average of $\Or(1/\sqrt{\epsilon})$ calls to $W(P)$, where $\epsilon$ is the probability of sampling an action from the stationary distribution. The algorithm achieving this is structured as follows. After an initialization stage where $\ket{\alpha^{\prime}}$ is prepared, a number of diffusion steps are carried out. Each such step consists of two parts. The first part is a reflection over the states corresponding to actions in the first copy of $\mathcal{H}$. In the second part, an approximate reflection over the state $\ket{\alpha^{\prime}}$, the mixing, is carried out~\cite{Paparo2014}. This second step involves $\Or(1/\sqrt{\delta})$ calls to $W(P)$.

The two-part diffusion steps are repeated $\Or(1/\sqrt{\epsilon})$ times before a sample is taken from the resulting state by measuring in the basis $\{\ket{i}\}_{i=1,\ldots,N}$. If an action is sampled, the algorithm concludes and that action is chosen as output. Otherwise, all steps are repeated. Since the algorithm amplifies the probability of sampling an action (almost) to unity, carrying out the deliberation procedure with the help of such a Szegedy walk hence requires an average of $\Or(1/\sqrt{\delta\epsilon})$ calls to $W(P)$. In comparison, a classical RPS agent would require an average of $\Or(1/\delta)$ applications of $P$ to mix the Markov chain, and an average of $\Or(1/\epsilon)$ samples to find an action. Q-RPS agents could hence achieve a quadratic speed-up in their reaction time.

Here, it should be noted that, its elegance not withstanding, the construction of the approximate reflection for general RPS networks is demanding for current quantum computational architectures. Most notably, this is due to the requirement of two copies of $\mathcal{H}$, on which frequently updated\footnote{Updating of the clip network may include, e.g., modifications of the weights associated to the edges of the graph corresponding to the clip network in such a way that weights of connections between percepts and rewarded actions are increased. In addition, updates may involve the addition or deletion of clips, as well as more sophisticated mechanisms such as glow, generalization, etc., see Refs. \cite{Mautner2015,Paparo2014,Melnikov2017}.} coherent conditional operations need to be carried out~\cite{Dunjko2015,Friis2014,Friis2015}. However, for the special case of rank-one Markov chains $P$, the entire chain can be represented on one copy of $\mathcal{H}$ by a single unitary $U_{\!P}=U_{i}\; \forall i$, since all columns of $P$ are identical. Conceptually, this simplification corresponds to a situation where each percept-specific clip network contains only actions and the Markov chain is mixed in one step ($\delta=1$). In such a case one uses flags to mark desired actions. Interestingly, these minor alterations also allow to establish a one-to-one correspondence with the hitting-based basic PS using two-layered networks, into which all standard tabular reinforcement learning models such as Q-learning or SARSA can be subsumed when the update, and transition rules have been appropriately amended~\cite{Russell2003}. In particular, basic PS using a two-layered network is already able to solve interesting classical tasks such as the mountain-car problem, grid-world, and many more~\cite{Mautner2015,Melnikov2014,Makmal2016,Hangl2016,Tiersch2015,Melnikov2018}.

Let us now discuss how the algorithm above can be performed for the rank-one case with the flagging mechanism in place. First, we restrict $\mathcal{A}$ to be the subspace of the flagged actions only, assuming that there are $n\ll N$ of these, and we denote the corresponding probabilities within the stationary distribution by $a_{1},\ldots,a_{n}$. In the initialization stage, the state $\ket{\alpha}=\sum_{i=1,\ldots,N} \sqrt{a_{i}}\ket{i}$ is prepared. Then, an optimal number of $k$ diffusion steps~\cite{Grover1996} is carried out, where
\begin{eqnarray}
	k &={\rm round} \left( \frac{\pi}{4\sqrt{\epsilon}} - \frac{1}{2} \right),
	 \label{eq:diff_step}
\end{eqnarray}
and $\epsilon=\sum_{i=1,\ldots,n}a_{i}$ is the probability to sample a flagged action from the stationary distribution. Within the diffusion steps, the reflections are performed only over all flagged actions, i.e.,
\begin{eqnarray}
{\rm ref}_{\mathcal{A}}=2\sum\limits_{i=1}^n \ket{i}\!\!\bra{i}-\mathds{1}.
\end{eqnarray}
In the rank-one case, the reflections over the stationary distribution $\alpha$ becomes an exact reflection
\begin{eqnarray}
{\rm ref}_{\alpha}&=2\ket{\alpha}\!\!\bra{\alpha}-\mathds{1}
\end{eqnarray}
and can be carried out on one copy of $\mathcal{H}$~\cite{Dunjko2015}. After the diffusion steps, a sample is taken and the agent checks if the obtained action is marked with a flag. If this is the case, the action is chosen as output, otherwise the algorithm starts anew.

While a classical RPS agents requires an average of $\Or(1/\epsilon)$ samples until obtaining a flagged action, this number reduces to $\Or(1/\sqrt{\epsilon})$ for Q-RPS agents. This quantum advantage is particularly pronounced when the overall number of actions is very large compared to $n$ and the environment is unfamiliar to the agent or has recently changed its rewarding pattern, in which case $\epsilon$ may be very small. Given some time, both agents learn to associate rewarded actions with a given percept, suitably add or remove flags, and adapt $P$ (and by extension $\pmb\alpha$). In the short run, however, classical agents may be slow to respond and the advantage of a Q-RPS agent becomes apparent. Despite the remarkable simplification of the algorithm for the rank-one case with flags, the quadratic speed-up is hence preserved\footnote{Since $\delta=1$, the speed-up is only possible, and is achieved for, the quantity $\epsilon$.}. 
{This simplification also leads to a reduction in experimental complexity, in terms of the required number of two-qubit gates.}


\section{Experimental Implementation of Rank-One RPS\label{sec:experiment}}

\subsection{Quantum Algorithm}

The proof-of-principle experiment that we report in this paper experimentally demonstrates the speed-up of quantum-enhanced learning agents. That is, we are able to empirically confirm both the quadratically improved scaling of $\Or(1/\sqrt{\epsilon})$, and the correct output according to the tail of the stationary distribution. Here, $\epsilon$ denotes the initial probability of finding a flagged action within the stationary distribution $\pmb\alpha = \{a_i\}$ for the average number of calls of the diffusion operator before sampling one of the desired actions.  The tail is defined as  the first $n$ components of $\pmb\alpha$. By a correct output according to the tail of the stationary distribution, we mean that $a_{j}/a_{k}=b_{j}/b_{k}\ \forall j,k\in\{1,\ldots,n\}$, where $b_j$ denotes the final probability that the agent obtains the flagged action labeled $j$. Note that the Q-RPS algorithm enhances the overall probability of obtaining a flagged action such that
\begin{eqnarray}
    \tilde{\epsilon}\,\equiv\,\sum\limits_{i=1}^{n}b_{i}   & >\,\sum\limits_{i=1}^{n}a_{i}\,=\,\epsilon,
    \label{eq:epsilon}
\end{eqnarray}
whilst maintaining the relative probabilities of the flagged actions according to the tail of $\pmb\alpha$, as illustrated in \fig{fig:la-concept}~(b).

For the implementation we hence need at least a three-dimensional Hilbert space that we realize in our experiment using two qubits encoded in the energy levels of two trapped ions (see the experimental setup section):
Two states to represent two different flagged actions (represented in our experiment by $\ket{00}$ and $\ket{01}$), and at least one additional state for all non-flagged actions ($\ket{10}$ and $\ket{11}$ in our experiment). The preparation of the stationary state is implemented by
\begin{eqnarray}
    \ket{\alpha}    &=\,U_{\!P}(\theta_{1},\theta_{2})\ket{00}\,=\,R_1(\theta_1,\frac{\pi}{2})R_2(\theta_2,\frac{\pi}{2})\ket{00},
\end{eqnarray}

where $R_j(\theta,\phi)$ is a single-qubit rotation on qubit $j$, i.e.,
\begin{eqnarray}
    R_j(\theta,\phi) &=\,\exp\bigl[\I \frac{\theta}{2}(X_j \cos \phi - Y_j \sin \phi)\bigr].
    \label{eq:rotation}
\end{eqnarray}
Here, $X_j$, $Y_j$, and $Z_{j}$ denote the Pauli operators of qubit $j$.
The total probability $\epsilon=a_{00}+a_{01}$ for a flagged action within the stationary distribution is then determined by $\theta_1$ via
\begin{eqnarray}
	\epsilon=\cos^2(\theta_1/2),
	\label{eq:theta1}
\end{eqnarray}
whereas $\theta_2$ determines the relative probabilities of obtaining one of the flagged actions via
\begin{eqnarray}
	a_{00}/\epsilon=\cos^2(\theta_2/2).
	\label{eq:theta2}
\end{eqnarray}

\begin{figure}[h]
    \centering
	\includegraphics[width=0.5\textwidth]{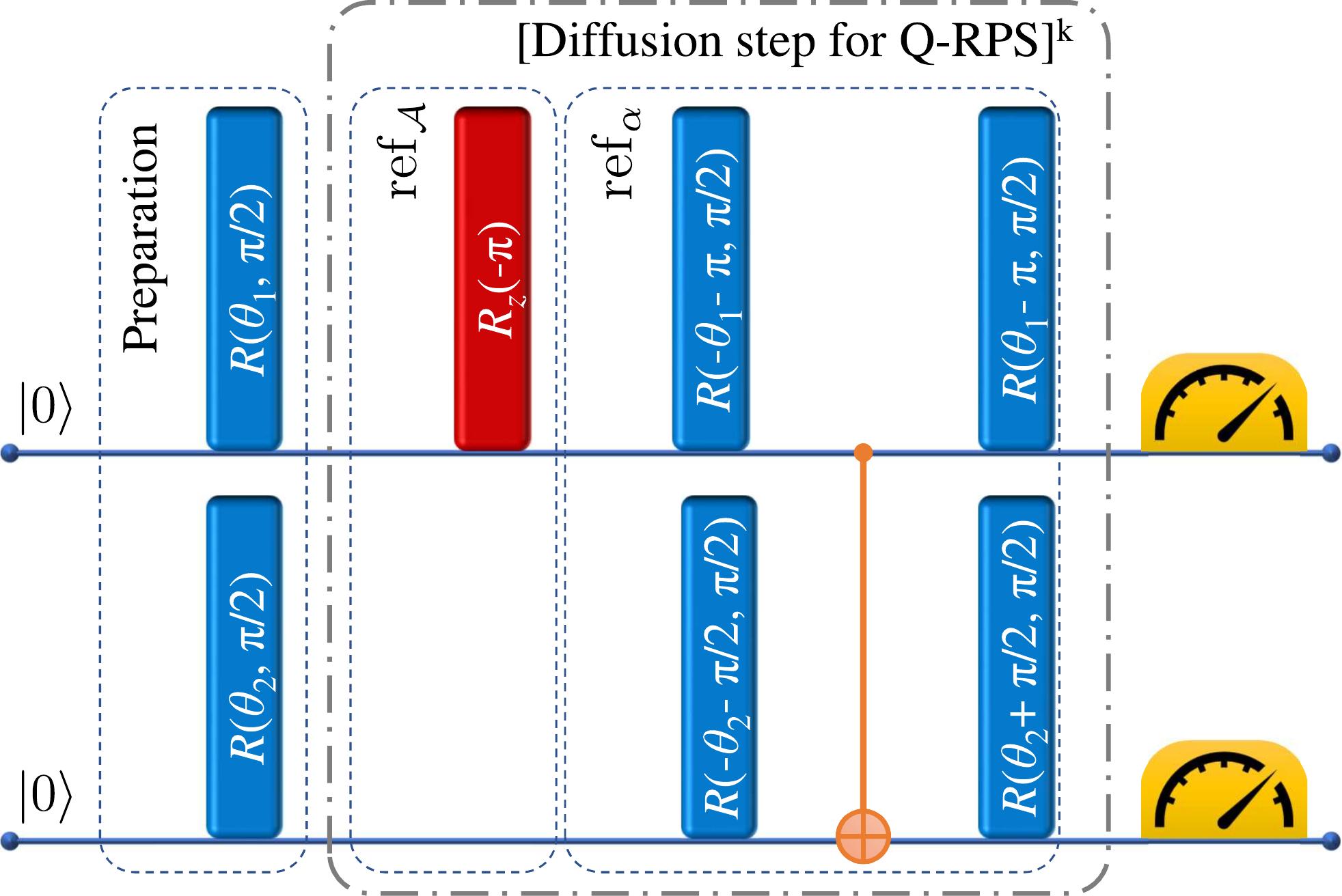}
    \caption{Quantum circuit for Q-RPS. A rank-one Q-RPS is implemented using two qubits. The diffusion step consisting of reflections over the flagged actions and the stationary distribution (shown
    once each) is repeated $k$ times, where $k$ is given by \eq{eq:diff_step} in \Sec{sec:theory part}. The specific pulse sequence implementing this circuit is explained in \Sec{sec:experimental setup}.}
	\label{fig:pulse-sequence-theo}
\end{figure}

The reflection over the flagged actions ${\rm ref}_\mathcal{A}$ is here given by a $Z$ rotation, defined by $R_{j,z}(\theta)=\exp[-\I \frac{\theta}{2}Z_j ]$, with rotation angle $-\pi$ for the first qubit,
\begin{eqnarray}
	{\rm ref}_{\mathcal{A}} = R_{1,z}(-\pi)=\exp[\I \frac{\pi}{2}Z_1 ].
\end{eqnarray}
The reflection over the stationary distribution can be performed by a combination of single-qubit rotations determined by $\theta_1$ and $\theta_2$ and a CNOT gate given by
\begin{eqnarray}
	\fl    
    {\rm ref}_{\alpha} &=
    R_{1}(\theta_{1}-\pi,\frac{\pi}{2})
    R_{2}(\theta_{2}+\frac{\pi}{2},\frac{\pi}{2})
    U\subtiny{0}{0}{\mathrm{CNOT}}
    R_{1}(-\theta_{1}-\pi,\frac{\pi}{2})
    R_{2}(-\theta_{2}-\frac{\pi}{2},\frac{\pi}{2}),
\end{eqnarray}
which can be understood as two calls to $U_{\!P}$ (once in terms of $U_{\!P}^{\dagger}$) supplemented by fixed single-qubit operations~\cite{Dunjko2015}. The total gate sequence for a single diffusion step (consisting of a reflection over the flagged actions followed by a reflection over the stationary distribution) can hence be decomposed into single-qubit rotations and CNOT gates and is shown in~\fig{fig:pulse-sequence-theo}. The speed-up of the rank-one Q-RPS algorithm {with respect to} a classical RPS agent manifests in terms of a quadratically smaller average number of calls to $U_{\!P}$ (or, equivalently, to the diffusion operator $D={\rm ref}_{\alpha}{\rm ref}_{\mathcal{A}}$) until a flagged action is sampled. Since the final probability of obtaining a desired action is $\tilde{\epsilon}\equiv\sum_{i=1,\ldots,n}b_{i}$, we require $1/\tilde{\epsilon}$ samples on average, each of which is preceded by the initial preparation of $\ket{\alpha}$ and $k$ diffusion steps. The average number of uses of $U_{\!P}$ to sample correctly is hence
\begin{eqnarray}
C=\bigl(2k(\epsilon)+1\bigr)/\tilde{\epsilon} \ ,
\end{eqnarray}
which we refer to as \emph{`cost'} in this paper. In what follows, it is this functional relationship between $C$ and $\epsilon$ that we put to the test, along with the predicted ratio $a_{00}/a_{01}$ of occurrence of the two flagged actions.

\subsection{The experimental setup\label{sec:experimental setup}}

Two $^{171}$Yb$^{+}$ ions are confined in a linear Paul trap with axial and radial trap frequencies of 2$\pi~\times~117$~kHz and $2\pi~\times~590$~kHz, respectively. After Doppler cooling, the two ions form a linear Coulomb crystal, which is exposed to a static magnetic field gradient of $19$~T/m, generated by a pair of permanent magnets. The ion-ion spacing in this configuration is approximately $10~\mu$m. Magnetic gradient induced coupling (MAGIC) between ions results in an adjustable qubit interaction mediated by the common vibrational modes of the Coulomb crystal~\cite{Khromova2012}. In addition, qubit resonances are individually shifted as a result of this gradient and become position dependent. This makes the qubits distinguishable and addressable by their frequency of resonant excitation. The addressing frequency separation for this two-ion system is about $3.7$~MHz. All coherent operations are performed using  radio frequency (RF) radiation near $12.6$~GHz, matching the respective qubit resonances~\cite{Piltz2014}. The RF power is carefully adjusted for each ion in order to achieve an equal Rabi frequency of $20.92(3)$~kHz. A more detailed description of the experimental setup is given in Refs.~\cite{Khromova2012,Woelk2015,Piltz2016}.

The qubits are encoded in the hyperfine manifold of each ion's ground state, representing an effective spin $1/2$ system. The qubit states $\ket{0}$ and $\ket{1}$ are represented by the energy levels $\ket{^2S_{1/2}\,,F=0\,}$ and $\ket{^2S_{1/2}\,,F=1, m_F=+1\,}$, respectively. The ions are Doppler cooled on the resonance  $\ket{^2S_{1/2}\,,F=1\,}$ $\leftrightarrow$ $\ket{^2P_{1/2}\,,F=0\,}$ with laser light near $369$~nm. Optical pumping into long-lived meta-stable states is prevented using  laser light near $935$~nm and $638$~nm. The vibrational excitation of the Doppler cooled ions is further reduced by employing RF sideband cooling for both the center of mass mode and the stretch mode~\cite{Sriarunothai2017}. This leads to a mean vibrational quantum number of $\langle n \rangle \le 5$ for both modes. The ions are then initialized in the qubit state $\ket{0}$ by state selective optical pumping with a $2.1$~GHz blue-shifted Doppler-cooling laser on the $\ket{^2S_{1/2}\,,F=1\,}$ $\leftrightarrow$ $\ket{^2P_{1/2}\,,F=1\,}$ resonance.


\subsection{State preparation, conditional dynamics, and read-out}

\begin{figure}[h]
	\centering
	\includegraphics[width=0.8\textwidth]{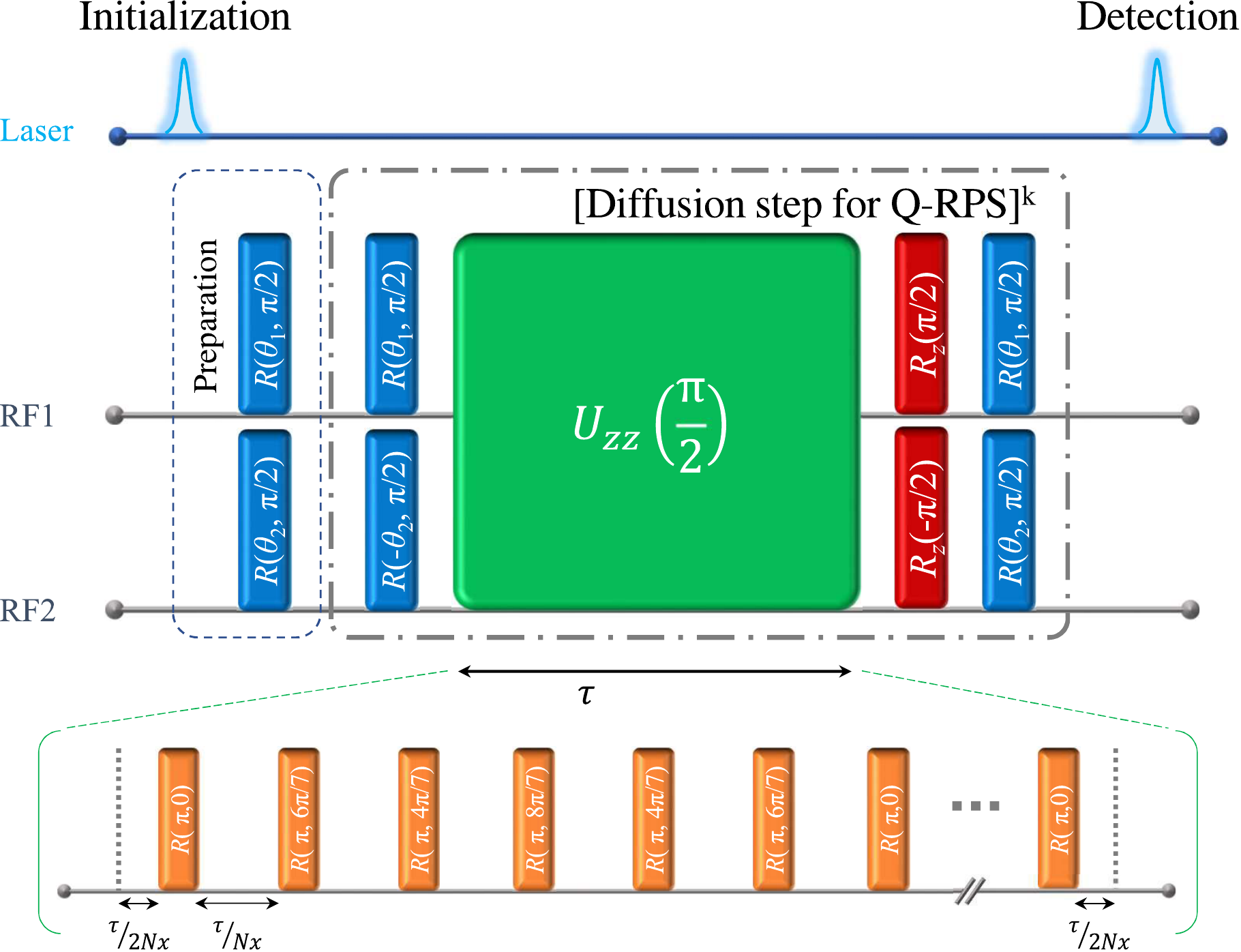}
	\caption{Experimental sequence for Q-RPS. RF1 and RF2 each indicate a time axis for a qubit. The qubits are prepared in the desired input states using single-qubit rotations implemented by applying RF pulses. For each RF pulse, the two parameters within the parentheses represent the specific rotation angle and phase according to \eq{eq:rotation} -- \eq{eq:theta2}. Also, dynamical decoupling (DD) during conditional evolution $U_{zz}(\pi/2)$ (indicated by a green box), is implemented using RF pulses (indicated in yellow). Ten sets of 14 pulses each (UR14)~\cite{Genov2017} are applied during the evolution time $\tau=4.24$~ms with a J-coupling between the two ions of $2\pi \times 59$ Hz. The diffusion step is repeated $k$ times according to \eq{eq:diff_step} in \Sec{sec:theory part}. Laser light near 369 nm is used for cooling and to initialize the ions in the qubit state $\ket{0} \equiv \ket{^2S_{1/2}\,,F=0\,}$. At the end of the coherent manipulation, laser light is used again for state selective detection and also for Doppler cooling. These process durations are: 30 ms for Doppler cooling, 100 ms for sideband cooling on the center-of-mass mode, 100 ms for sideband cooling on the stretch mode, 0.25 ms for initialization in state $\ket{0}$ of the ions, and 2 ms for detection.
	}
\label{fig:pulse-sequence}
\end{figure}

The desired qubit states are prepared by applying an RF pulse resulting in a coherent qubit rotation with precisely defined rotation angle and phase as given by 
\eq{eq:rotation} -- \eq{eq:theta2}.
We replace $R_{z}(\frac{\pi}{2})$  by $R(\frac{\pi}{2} , \frac{\pi}{2}) R(\frac{\pi}{2} , 0) R(\frac{\pi}{2} , \frac{3\pi}{2})$ and
$R_{z}(-\frac{\pi}{2})$  by $R(\frac{\pi}{2} , \frac{\pi}{2}) R(\frac{\pi}{2} , \pi) R(\frac{\pi}{2} , \frac{3\pi}{2})$. The required number of diffusion steps is then applied to both qubits, using appropriate single-qubit rotations and a two-qubit ZZ-interaction given by
\begin{eqnarray}
    U_{ZZ}(\theta)=\exp{[\I \frac{\theta}{2}Z_{1}Z_{2}]},
\end{eqnarray}
which is directly realizable with MAGIC~\cite{Khromova2012}. A CNOT gate ($U_\mathrm{CNOT}$) can then be performed via
\begin{eqnarray}
    U\subtiny{0}{0}{\mathrm{CNOT}}  &=
    e^{-i\frac{\pi}{4}}
    R_2(\frac{\pi}{2},\!\frac{3\pi}{2})U_{\!Z\hspace*{-0.5pt}Z}(\frac{\pi}{2})R_2(\frac{\pi}{2},0)R_{2,z}(\frac{\pi}{2})R_{1,z}(-\frac{\pi}{2}).
    \nonumber
\end{eqnarray}
The required number of single qubit gates is optimized by combining appropriate single qubit rotations together from ${\rm ref}_\mathcal{A}$ and ${\rm ref}_\alpha$ (see \fig{fig:pulse-sequence-theo}). Thus, we can simplify the algorithm to
\begin{eqnarray}
	\fl
	D =& R_2(\theta_2,\frac{\pi}{2}) R_1(\theta_1,\frac{\pi}{2})
	 	 R_{2,z}(-\frac{\pi}{2}) R_{1,z}(\frac{\pi}{2})
	     U_{ZZ}(\frac{\pi}{2}) R_2(-\theta_2,\frac{\pi}{2}) R_1(\theta_1,\frac{\pi}{2})   ,
\end{eqnarray}
as shown in~\fig{fig:pulse-sequence}.

During the evolution time of $4.24$~ms for $U_{ZZ}$ in each diffusion step both qubits are protected from decoherence by applying universally robust (UR) dynamical decoupling (DD) pulses~\cite{Genov2017}. A set of ten ($x=10$) UR14 ($N=14$) RF $\pi$-pulses, equaling a total of 140 pulses, is applied. Each set is comprised of 14 error cancelling pulses (\fig{fig:pulse-sequence}) with appropriately chosen phase $\phi$:
\begin{eqnarray*}
	\left( 0, \frac{6\pi}{7}, \frac{4\pi}{7}, \frac{8\pi}{7}, \frac{4\pi}{7}, \frac{6\pi}{7}, 0,0, \frac{6\pi}{7}, \frac{4\pi}{7}, \frac{8\pi}{7}, \frac{4\pi}{7}, \frac{6\pi}{7}, 0\right).
\end{eqnarray*}
\noindent
Since the phases of the $\pi$-pulses are symmetrically arranged in time, only the first seven pulses are shown in~\fig{fig:pulse-sequence}. The last pulse is also shown to visualize the spacing of these pulses with respect to the start and end of evolution time, compared to the intermediate pulses.
The maximum interaction time of 30 ms required to realize the deliberation algorithm (corresponding to 7 diffusion steps) is 60 times longer than the qubit coherence time. Such a long coherent interaction time is accomplished by the DD pulses applied to each qubit simultaneously.

Finally, projective measurements on both qubits are performed in the computational basis $\{\ket{0},\ket{1}\}$ by scattering laser light near $369$ nm on the $\ket{^2S_{1/2}\,,F=1\,}$ $\leftrightarrow$ $\ket{^2P_{1/2}\,,F=0\,}$ transition, and detecting spatially resolved resonance fluorescence using an electron multiplying charge coupled device to determine the relative frequencies  $b_{00},b_{01},b_{10}, b_{11}$ for obtaining the states $\ket{00}$, $\ket{01}$, $\ket{10}$, and $\ket{11}$, respectively. Two thresholds are used to distinguish between dark and bright states of the ions, thus discarding 10\% of all measurements as ambiguous events with a photon count that lies in the region of two partially overlapping Poissonian distributions representing the dark and bright states of the ions~\cite{Vitanov2015,Woelk2015}.



\section{Experimental Results\label{sec:results}}

As discussed above, our goal is to test the two characteristic features of rank-one Q-RPS: (i) the scaling of the average cost $C$ with $\epsilon$, and (ii) the sampling ratio for the different flagged actions. 

Therefore, our first set of measurements studies the behavior of the cost $C$ as a function of the total initial probability $\epsilon$. The second set of measurements studies the behavior of the output probability ratio $r_{f}=b_{00}/b_{01}$ as a function of input probability ratio $r_{i}=a_{00}/a_{01}$.

\begin{figure}[h]
	\centering
	\includegraphics[width=0.5\textwidth]{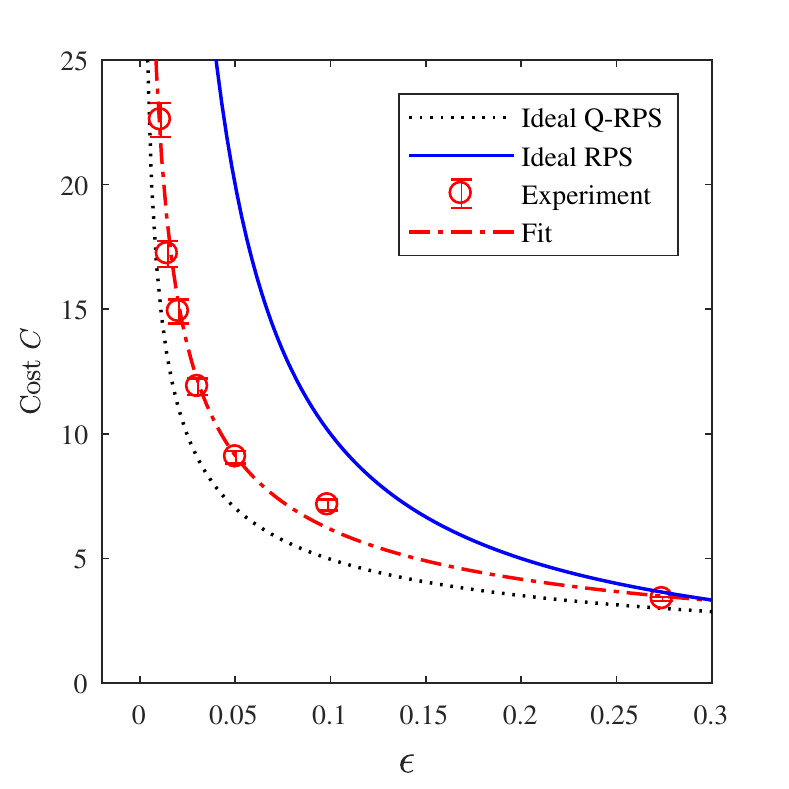}
	\caption{Scaling behavior of the learning agent's cost employing Q-RPS and RPS. After the preparation of $\ket{\alpha}$, $k$ diffusion steps are applied before an action is sampled. This procedure is repeated until a flagged action is obtained, accumulating a certain cost $C$, whose average is shown on the vertical axis. Measurements are performed for different values of $\epsilon$ corresponding to $k=1$ to $k=7$ diffusion steps. The dashed black line and the solid blue line represent the behavior expected for ideal Q-RPS ($\epsilon^{-0.5}$) and ideal classical RPS ($\epsilon^{-1}$), respectively. The fit to the experimental data confirms that the scaling behavior follows a $\epsilon^{-0.57}$ behavior, and thus is consistent with Q-RPS. The data show that the experimental Q-RPS outperforms the classical RPS within the range of $\epsilon$ chosen in the experiment. Error bars represent the statistical errors.
	}
	\label{fig:resource}
\end{figure}

For the former, a series of measurements is performed for different values of $\epsilon$ corresponding to $k=1$ to $k=7$ diffusion steps after the initial state preparation (table 1). To obtain the cost $C=\bigl(2k(\epsilon)+1\bigr)/\tilde{\epsilon}$, where $\tilde{\epsilon}=b_{00}+b_{01}$, we measure the probabilities $b_{00}$ and $b_{01}$ after $k$ diffusion steps and repeat the experiment 1600 times for fixed $\epsilon$. The average cost is then plotted against $\epsilon$ as shown in~\fig{fig:resource}. The algorithm complexity is defined as the number of computational steps (equivalently, the number of calls to $U_P$) until the flagged action is sampled. To describe the algorithm complexity, the number of operations can be expressed as $\Or(\epsilon^{- \xi})$. Ideally, the RPS gives $\xi = 1$ whereas the Q-RPS gives  $\xi = 0.5$.
The experimental data shows that the cost decreases with $\epsilon$ where $\xi = 0.57(5)$. This is in good agreement with the behavior expected for the ideal Q-RPS algorithm. In the range of chosen probabilities $\epsilon$, the experimental result of Q-RPS shows improved scaling as compared to the expected classical RPS, and clearly outperforms the classical RPS, as shown in \fig{fig:resource}. The deviation from the ideal behavior is attributed to a small detuning of the RF pulses implementing coherent operations, as we discuss in \Sec{sec:error-discussion}.

\begin{figure}[h]
	\centering
	\includegraphics[width=0.5\textwidth]{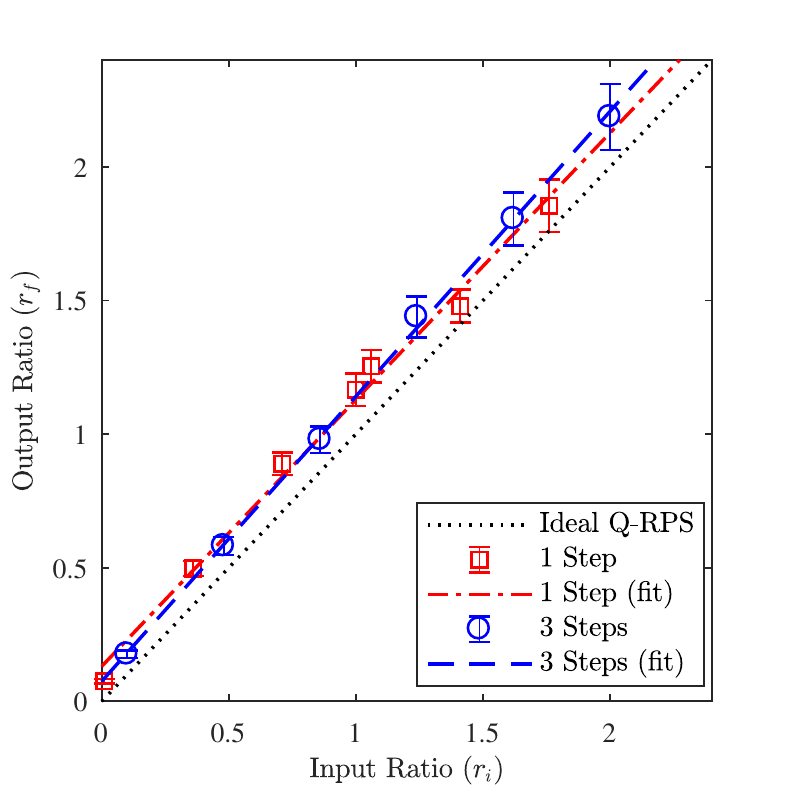}
	\caption{Output distribution. A comparison of the output ratio of two flagged actions at the end of the algorithm with the corresponding input ratio is shown. Measurements are performed with $k=1$ (red square) and $k=3$ (blue circle) diffusion steps. The black dashed line shows the behavior of the ideal Q-RPS. The red and blue dashed lines, each representing a linear fit to the corresponding set of data, confirm that the initial probability ratio is maintained. Error bars represent statistical errors.}
	\label{fig:ratio}
\end{figure}

For the second set of measurements, we select calculated probabilities $a_{00}$ and $a_{01}$ in order to obtain different values of the input ratio $r_{i}=a_{00}/a_{01}$ between 0 and 2, whilst keeping $k(\epsilon)$ in a range between $k=1$ and $k=3$ (table 2).
For these probabilities $a_{00}$ and $a_{01}$, the corresponding rotation angles $\theta_1$ and $\theta_2$ of RF pulses used for preparation are extracted using \eq{eq:theta1} and \eq{eq:theta2}.
We then carry out the Q-RPS algorithm for the specific choices of $k$ and repeat it 1600 times to estimate the probabilities $b_{00}$ and $b_{01}$. We finally obtain the output ratio $r_{f}=b_{00}/b_{01}$, which is plotted against the input ratio in~\fig{fig:ratio}. The experimental data follows a straight line with a small offset from the ideal behavior $r_f/r_i=1$. Therefore, the ratio of the number of occurrences of the two actions obtained at the end of the deliberation process is maintained with respect to the relative probabilities of the initial stationary distribution.

The slopes of the two fitted linear functions shown in~\fig{fig:ratio} agree within their respective error showing that the deviation of the output ratio from the ideal result is independent of the number of diffusion steps.  In addition, this indicates that this deviation is not caused  by the quantum algorithm itself, but by the initial state preparation and/or by the final measurement process where such a deviation can be caused by an asymmetry in the detection fidelity 
(see \Sec{sec:error-discussion}). 
Indeed, the observed deviation is well explained by a typical asymmetry in the detection fidelity of 3\% as encountered in the measurements presented here. This implies reliability of the quantum algorithm also for a larger number of diffusion steps.


\subsection{Interpretational considerations\label{sec:error-discussion}}

In this section, we discuss deviations of the experimental data from idealized theory predictions. In particular, for the chosen values of $\epsilon$ and the corresponding optimal $k(\epsilon)$, it is expected that the probability of obtaining a flagged action is close to $100\%$. However, the success probability in our experiment lies between $66\%$ (for $k=7$) and $88\%$ (for $k=1$). In what follows, we  discuss several reasons for this. First, we consider in detail experimental imperfections that affect the scaling of cost $C$ with $\epsilon$ as shown in \fig{fig:resource}. Then, we discuss how the input and output ratios (\fig{fig:ratio}), $r_i$ and $r_f$, are affected by an imbalanced detection efficiency for both qubit states.  In both cases the observed deviations from the ideal results are quantitively explained by numerically simulating the quantum algorithm taking into account experimental imperfections.

\subsubsection{Scaling of cost $C$.}

Even in an ideal scenario without noise or experimental imperfections the success probability $\tilde{\epsilon}$, as defined in \eq{eq:epsilon}, after $k$ diffusion steps is usually not equal to unity, and depends on the specific value of $\epsilon$. This behavior originates from the step-wise increase of the number of diffusion steps 
$k={\rm round}({\pi}/{(4\sqrt{\epsilon})}-\frac{1}{2})$ 
in the algorithm. The success probability is hence only $100\%$, if $k$ is an integer without rounding. The change of the ideal success probability with deviations of $\epsilon$ from such specific values is largest for small numbers of diffusion steps (e.g., $k=1$) and can drop down to $82\%$ (neglecting the cases where it is not advantageous to use a quantum algorithm at all). For larger numbers of diffusion steps, the exact value of $\epsilon$ does not play an important role any more for the ideal success probability provided that the correct number of diffusion steps is chosen. For example, for $k=6$, the ideal success probability is larger than $98\%$ independently of the exact value of $\epsilon$. Throughout this paper, we have chosen $\epsilon$ in such a way, that 
$({\pi}/{(4\sqrt{\epsilon})}-\frac{1}{2})$ 
(see \eq{eq:diff_step}) is always close to an integer (see \tab{tab:epsilon}), such that the deviation from a $100\%$ success probability due to the theoretically chosen $\epsilon$ is negligible compared to other error sources.

However, in a real experiment, the initial state, and therefore $\epsilon$, can only be prepared with a certain accuracy. This can lead to an inaccurate estimation of the optimal number of diffusion steps. As opposed to the ideal case, an assumed accuracy of $\epsilon \pm 1\%$ for the preparation  only has a small effect on the success probability $\tilde{\epsilon}$ (drop of less than $5\%$) for $\epsilon\gg 0.01$, corresponding to $k\leq 3$.  However, when $\epsilon$ does not fulfil the aforementioned condition and approaches $\approx 0.01$ from above, corresponding to  $k=6$, then the success probability drops down to $\tilde{\epsilon}=70\%$ due to a non-optimal choice of $k$.

The preparation accuracy depends on the detuning $\Delta\omega$ of the RF pulses for single-qubit rotations as well as on the uncertainty $\Delta \Omega$ in the determination of the Rabi frequency $\Omega$. The calibration of our experiment revealed $\Delta\omega/\Omega<0.05$ and $\Delta \Omega/\Omega=0.0015$ leading to an error in $ \epsilon$ of $\pm 2.5\cdot 10^{-3}$ and a decrease of the success probability $\tilde{\epsilon}$ of less than $0.04$. The detuning $\Delta\omega$ and the uncertainty of the Rabi frequency $\Delta\Omega$ not only influence the state preparation at the beginning of the quantum algorithm, but also its  fidelity, as is detailed in the next paragraph.

\begin{figure}[h]
	\centering
	\includegraphics[width=0.5\textwidth]{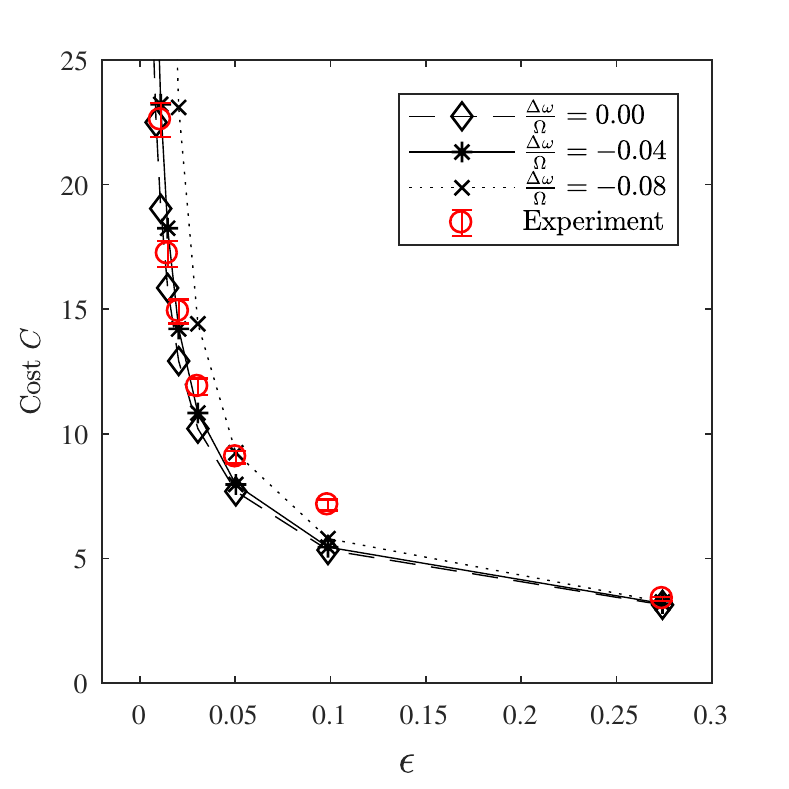}
	\caption{
	Detuning affecting scaling cost $C$. The influence of the detuning  of the RF pulses on the fidelity of the Q-RPS algorithm is shown for three different values of the relative detuning $\Delta\omega/\Omega$. Black markers indicate the results of numerical simulations of the complete Q-RPS algorithm taking different values of the relative detuning $\Delta\omega/\Omega$ into account. Most of the experimental data (red circles) lie close to a relative detuning of \protect$\Delta\omega/\Omega=-0.04$.
	\label{fig:scaling_error}}
\end{figure}

To prevent decoherence during conditional evolution, we use $140$~RF $\pi$-pulses per diffusion step and ion. Therefore, already a small detuning influences the fidelity of the algorithm. Consequently, the error induced by  detuning is identified as the main error source leading, for example, to $\tilde{\epsilon}\approx 0.77$ for $k=6$ and $\Delta\omega/\Omega = -0.04$. This error is much larger than the error caused by dephasing (that is still present after DD is applied), or the detection error. In a separate measurement, we determined an exponential dephasing rate of $\gamma \tau\approx 1/14$ for a single diffusion step of duration $\tau\approx 4$ ms, which would lead to $\tilde{\epsilon}\approx 0.90$ for $k=6$. Here, $\gamma$ indicates the experimentally diagnosed rate of dephasing, and  $\tau$ is the time of coherent evolution. The influence of the detuning  on the cost of our algorithm is shown in~\fig{fig:scaling_error} for different detunings. Here, we simulated  the complete quantum algorithm including the experimentally determined dephasing and detection errors for $\Delta\omega/\Omega \in \{0, -0.04, -0.08\}$. The experimental data is consistent with an average negative detuning of $\Delta\omega/\Omega = -0.04$. Note that the detuning not only influences the single-qubit rotations that are an integral part of the quantum algorithm, but also leads to errors during the conditional evolution when dynamical decoupling pulses are applied.

\subsubsection{Input and output ratios.}

In the ideal algorithm, the output ratio $r_f=b_{00}/b_{01}$ of the two flagged actions represented by the states $\ket{00}$ and $\ket{01}$ at the end of the algorithm equals the input ratio $r_i$. However, in the experiment we have observed deviations from $r_f/r_i=1$. During the measurements for the investigation of the scaling behavior (\fig{fig:resource}), we fixed $r_i=1$. The observed output ratios are varying by $0.98\leq r_f/r_i\leq 1.33$. That is, the probability $b_{00}$ to obtain the state $\ket{00}$ is increased {with respect to} $b_{01}$. Also during the measurement testing the output ratio, we observe that the output ratios are larger than the input ratios.

\begin{figure}[h]
	\centering
	\includegraphics[width=0.5\textwidth]{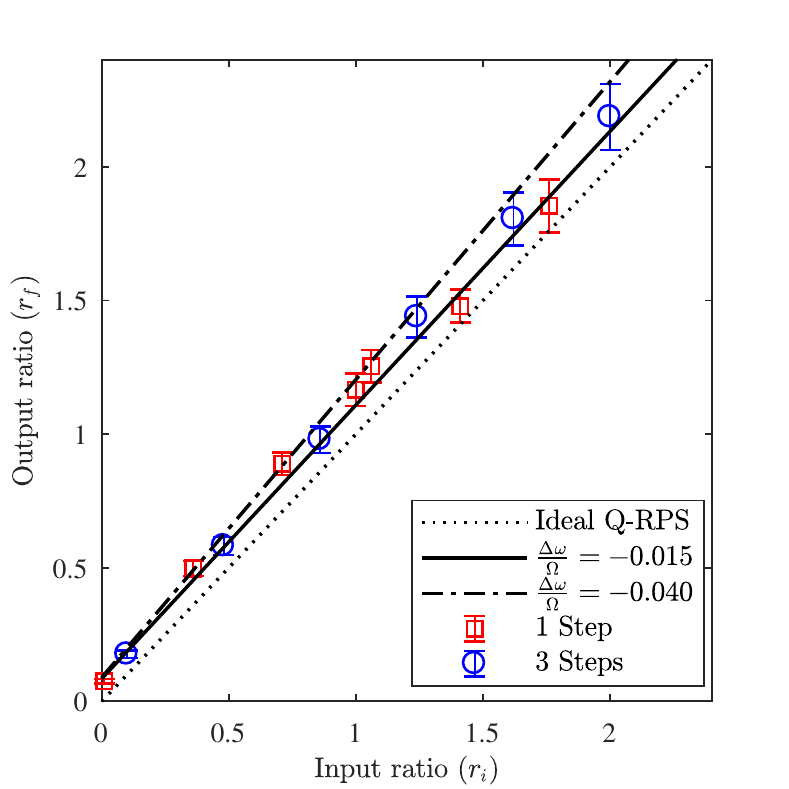}
	\caption{Imbalanced detection. The measured values (red squares and blue circles) of the input and output ratios are compared to simulations (solid and dash-dot black lines) of the Q-RPS algorithm taking into account the experimentally determined detection error and detuning errors. The solid line corresponds to an expected output ratio taking into account an unbalanced detection error where $d_{\mathrm{B}}=0.06$ for bright ions and $d_{\mathrm{D}}=0.03$ for dark ions and detuning error $\Delta\omega/\Omega= -0.04$. The dash-dot line represents the same detection error and a detuning error of $-$0.015.}
	\label{fig:ratio_error}
\end{figure}

An asymmetric detection error could be the cause for this observation. Typical errors in our experiment are given by the probability to detect a bright ion ($\ket{1}$) with a probability of $d_{\mathrm{B}}=0.06$  as dark, and a dark ion ($\ket{0}$) with a probability of $d_{\mathrm{D}}=0.03$ as bright. In~\fig{fig:ratio_error} we compare the measured output ratios with the calculated output ratios assuming the above mentioned detection errors and two different detuning errors, $\Delta\omega/\Omega \in \{-0.015, -0.04\}$.

When the experimentally determined detection error is taken into account, the simulation with detuning error $\Delta\omega/\Omega = -0.04$ does not describe the experimental data well for both one diffusion step and three diffusion steps. The experimental data agree well with a simulation using an average detuning error $\Delta\omega/\Omega = -0.015$. This indicates that the detuning during these measurements was kept around $\Delta \omega/\Omega = -0.015$ leading to an average success probability of $\tilde{\epsilon}=85\%$ for $k=3$ diffusion steps compared to $\tilde{\epsilon}=77\%$ for $k=3$ during the measurements investigating the scaling (see \tab{tab:epsilon}). In addition, errors in the preparation of the input states play a role, especially when preparing very large or very small ratios leading to either $a_{00}$ or $a_{01}$ being close to the preparation accuracy of $\leq 2.5\cdot 10^{-3}$.



\begin{table}[h]
\caption{Experimentally realized success probabilities. Initial theoretical probabilities, $\epsilon$, of finding a flagged action within the stationary distribution for various diffusion steps are shown. Success probabilities ($\tilde{\epsilon}$), that are theoretically calculated and experimentally measured, for diffusion steps $k$ = 1-7 are also shown.
	\label{tab:epsilon}}
\begin{tabular*}{1\textwidth}{c | @{\extracolsep{\fill}} c c c | c c c| c c c }
\hline
 &  & Theory &  &  & Theory &  & & Experiment &  \\
$k$ & $a_{00}$ & $a_{01}$ & $\epsilon$ & $b_{00}$ & $b_{01}$ & $\tilde{\epsilon}$& $b_{00}$ & $b_{01}$ & $\tilde{\epsilon}$ \\
\hline
1 &	0.1371 &	0.1371 &	0.2742 &	0.4966	& 0.4966	& 0.9932  &	0.449(15) &	0.440(15) &	0.89(2)\\

2 &	0.0493 &	0.0493 &	0.0987 &	0.4996	& 	0.4996	& 	0.9993 &	0.347(15) &	0.353(15) &	0.70(2)\\

3 &	0.0252 &	0.0252 &	0.0504 &	0.4999	& 	0.4999	& 	0.9998 &	0.438(16) &	0.334(15) &	0.77(2)\\

4 &	0.0152 &	0.0152 &	0.0305 &	0.5000 & 0.5000 & 1.0000&	0.422(15) &	0.336(15) &	0.76(2) \\

5 &	0.0102 &	0.0102 &	0.0204 &	0.5000 & 0.5000 & 1.0000&	0.407(17) &	0.331(16) &	0.74(2) \\

6 &	0.0073 &	0.0073 &	0.0146 &	0.5000 & 0.5000 & 1.0000&	0.431(17) &	0.324(16) &	0.76(2) \\

7 &	0.0055 &	0.0055 &	0.0110 &	0.5000 & 0.5000 & 1.0000 &	0.365(15) &	0.299(14) &	0.66(2) \\
\hline

\end{tabular*}
\label{tab:scaling}
\end{table}

\begin{table}[h]
\caption{Input and output distributions. Input and output ratios, $r_i$ and $r_f$ respectively, of the two flagged actions represented by the states $\ket{00}$ and  $\ket{01}$ for diffusion steps $k$ = 1 and $k$ = 3 are shown.
		}
\begin{tabular*}{1\textwidth}{c | @{\extracolsep{\fill}} c c c | c c c }
\hline
 &  & Theory &  &  & Experiment &  \\
$k$ & $a_{00}$ & $a_{01}$ & $r_i$ & $b_{00}$ & $b_{01}$ & $ r_f $ \\
\hline
1	&	0.00271	&	0.27144	&	0.01	&	0.061(7)	&	0.809(12)	&	0.075(9) \\
1	&	0.07257	&	0.20159	&	0.36	&	0.290(14)	&	0.583(15)	&	0.50(3) \\
1	&	0.11383	&	0.16032	&	0.71	&	0.415(15)	&	0.466(15)	&	0.89(4) \\
1	&	0.14107	&	0.13309	&	1.06	&	0.488(15)	&	0.389(15)	&	1.25(6) \\
1	&	0.16040	&	0.11376	&	1.41	&	0.519(13)	&	0.351(12)	&	1.48(6) \\
1	&	0.17482	&	0.09933	&	1.76	&	0.566(15)	&	0.305(14)	&	1.85(10) \\
1	&	0.13708	&	0.13708	&	1.00	&	0.468(16)	&	0.401(16)	&	1.17(6) \\
\hline
3	&	0.00458	&	0.04578	&	0.10	&	0.127(10)	&	0.718(14)	&	0.176(14) \\
3	&	0.01633	&	0.03402	&	0.48	&	0.301(15)	&	0.518(16)	&	0.58(3) \\
3	&	0.02328	&	0.02707	&	0.86	&	0.442(16)	&	0.451(16)	&	0.98(5) \\
3	&	0.02788	&	0.02248	&	1.24	&	0.510(16)	&	0.354(15)	&	1.44(8) \\
3	&	0.03114	&	0.01922	&	1.62	&	0.551(16)	&	0.305(14)	&	1.81(10) \\
3	&	0.03357	&	0.01679	&	2.00	&	0.586(15)	&	0.268(13)	&	2.19(12) \\
\hline
\end{tabular*}
\label{tab:ratio}
\end{table}



\section{Conclusion\label{sec:conclusion}}

We have investigated a quantum-enhanced deliberation process of a learning agent implemented in an ion trap quantum processor. Our approach is centered on the projective simulation~\cite{Briegel2012} model for reinforcement learning. Within this paradigm, the decision-making procedure is cast as a stochastic diffusion process, that is, a (classical or quantum) random walk in a representation of the agent's memory.

The classical PS framework can be used to solve standard textbook problems in reinforcement learning~\cite{Mautner2015,Melnikov2014,Makmal2016}, and has recently been applied in advanced robotics~\cite{Hangl2016}, adaptive quantum computation~\cite{Tiersch2015}, as well as in the machine-generated design of quantum experiments~\cite{Melnikov2018}. We have focused on reflecting projective simulation~\cite{Paparo2014}, an advanced variant of the PS model based on ``mixing", where the deliberation process allows for a quantum speed-up of Q-RPS agents {with respect} to their classical counterparts. In particular, we have considered the interesting special case of rank-one Q-RPS. This provides the advantage of the speed-up offered by the mixing-based approach, but is also in one-to-one correspondence with the hitting-based basic PS using two-layered networks, which has been applied in classical task environments~\cite{Mautner2015,Melnikov2014,Makmal2016,Hangl2016,Tiersch2015,Melnikov2018}. In addition, rank-one Q-RPS can be used to encode all tabular reinforcement learning models including Q-Learning and SARSA by appropriately amending the update and transition rules~\cite{Russell2003}.

In a proof-of-principle experimental demonstration, we verify that the deliberation process of the quantum learning agent is quadratically faster compared to that of a classical learning agent. The experimental uncertainties in the reported results, which are in excellent agreement with a detailed model, do not interfere with this genuine quantum advantage in the agent's deliberation time. We achieve results for the cost $C$ for up to 7 diffusion steps corresponding to an initial probability $\epsilon$ = 0.01 to choose a flagged action. In this sense, our experimental realization of a {rank-one Q-RPS} decision-making algorithm, which differs from standard amplitude amplification already due to the reflection over the stationary (rather than a uniform) distribution, also provides a comprehensive test of the scaling behaviour that goes beyond previous experiments~\cite{Brickman2005,DiCarlo2009,Figgatt2017}, where standard amplitude amplification based on single diffusion steps has been carried out.

The systematic variation of the ratio $r_i$ between the input probabilities, $a_{00}$ and $a_{01}$ for flagged actions and the measurement of the ratio $r_f$ between the learning agent's output probabilities, $b_{00}$ and $b_{01}$ as a function of $r_i$ shows that the quantum algorithm is reliable independent of the number of diffusion steps.

This experiment highlights the potential of a quantum computer in the field of quantum enhanced learning and artificial intelligence. A practical advantage, of course, will become evident once larger percept spaces and general rank-$N$ Q-RPS are employed.
Such extensions are, from the theory side, unproblematic given that the modularized nature of the algorithm makes it scalable{, following \citen{Dunjko2015}}. 
An experimental realization of such large-scale quantum enhanced learning will be feasible with the implementation of scalable quantum computer architectures.  Meanwhile, all essential elements of Q-RPS have been successfully demonstrated in the proof-of-principle experiment reported here.


\ack{
S.~W. thanks A. Melnikov for fruitful discussions. T.~S., S.~W., G.~S.~G. and C.~W. acknowledge funding from Deutsche Forschungsgemeinschaft. G.~S.~G. also acknowledges support from the European Commission's Horizon 2020 research and innovation program under the Marie Sk\l{}odowska-Curie grant agreement number 657261. H.~J.~B. acknowledges support from the Austrian Science Fund (FWF) through the Grant No. SFB FoQuS F4012. N.~F acknowledges support from the Austrian Science Fund (FWF) through the project P 31339-N27, the START project Y879-N27, and the joint Czech-Austrian project MultiQUEST (I 3053-N27 and GF17-33780L).
}


\section*{References}
\bibliographystyle{iopart-num.bst}
\bibliography{bibfile}


%
%

\end{document}